\documentstyle[prb,aps,multicol,epsfig]{revtex}

\begin{document} \bibliographystyle{unsrt} 
\input{epsf}  
\title{Sensitivity to temperature perturbations of the ageing states 
in a re-entrant ferromagnet}
\author{K. Jonason and P. Nordblad}

\address{Department of Materials Science, Uppsala University,
Box
534, \\ S-751 21 Uppsala, Sweden }
\maketitle

\begin{abstract}

Dynamic magnetic properties and ageing phenomena of the re-entrant 
ferromagnet (Fe$_{0.20}$Ni$_{0.80}$)$_{75}$P$_{16}$B$_{6}$Al$_{3}$ are 
investigated by time dependent
zero field cooled magnetic relaxation, $m (t)$, measurements. 
The influence of a temperature cycling (perturbation), $\pm$  $\Delta T$,
(prior the field application) on the relaxation rate   is investigated 
both in the low temperature re-entrant spin glass 'phase' and in the
ferromagnetic phase. In the ferromagnetic phase the influence of a positive
and a negative temperature cycle (of equal magnitude) on the response is
almost the same (symmetric response). The result at lower temperatures, 
in the RSG 'phase' is asymmetric, with  a strongly affected response for
positive, and hardly no influence on the response for negative temperature cycles. 
The behaviour at low temperatures is similar to what is observed in ordinary 
spin glasses.

\end{abstract}
\vskip 0.2cm

\centerline{PACS numbers: 75.40.Gb 75.50.Lk}
\vskip 0.2cm

\bigskip
\begin{multicols}{2}
\narrowtext

\section {Introduction}

The field of random magnets provides a vast area of challenging problems. One problem
 of great interest is re-entrant magnets where a competition between spin glass and 
 ferro- or antiferromagnetic  long range order is present. When the temperature is 
 lowered in such a system, there is a transition from the paramagnetic to a ferro- 
 or antiferromagnetic phase and when further lowering the temperature a transition 
 to a re-entrant spin glass 'phase' might take place. This behaviour has been seen 
 in a number of disordered magnetic materials, both ferromagnetic and 
 antiferromagnetic, and it also occurs in mean field theory of random magnets 
 \cite{one}. 
	
	True re-entrance occurs in a re-entrant ferromagnet if the long range 
	ferromagnetic order eventually disappears at a finite temperature 
	$T_{RSG}$ 
	and is succeeded by a new equilibrium phase, the re-entrant spin glass phase. 
	The problem with spin glasses is that they never reach equilibrium on laboratory
	 (or even geological) time scales thus when the low temperature 'phase' in a 
	 re-entrant system is of spin glass character, the interpretation of the 
	 experimental results becomes hazardous. In previous papers we have reported 
	 results from magnetisation and susceptibility measurements on the dynamics 
	 of the re-entrant ferromagnet (Fe$_{0.20}$Ni$_{0.80}$)$_{75}$P$_{16}$B$_{6}$Al$_{3}$
	 \cite{two,three,four}. Similarities 
	 but also distinct differences between the dynamics in the RSG and the FM 
	 regions have been found, e.g. ageing is found in both the FM and the RSG 
	 'phase' \cite{two,three} but an established ageing state is destroyed by much weaker 
	 magnetic field perturbations in the FM than in the RSG region 
	 \cite{four}.
	
	In this paper we report striking differences in how an established ageing  
	state is affected by temperature cyclings (perturbations) in the RSG 
	ÔphaseÕ as compared to in the FM phase. Results which can be interpreted 
	to support the development of a true spin glass phase at low temperatures.

\section{Experimental}

In slowly relaxing magnetic systems (e.g. spin glasses, disordered ferromagnets etc. ),
 a non equilibrium (ageing) behaviour may be revealed from dc-magnetic relaxation experiments. 
 In such an experiment the sample is cooled from a high temperature, 
 $T_{ref}$, in the paramagnetic
  phase, to the measurement temperature, $T_m$, where it is kept a certain 
  wait time, $t_{w}$. Thereafter
   a small magnetic field, $h$,  is applied and the magnetic relaxation, 
   $m (t)$, is recorded as a
    function of time. When this response is dependent on the wait time, the system displays a 
    non equilibrium behaviour, i.e.  it ages. In this investigation we study how the magnetic 
    response changes when a temperature cycling, $\Delta T$, is made prior to the field application  
    both in the FM phase and the RSG ÔphaseÕ of the re-entrant ferromagnet 
    (Fe$_{0.20}$Ni$_{0.80}$)$_{75}$P$_{16}$B$_{6}$Al$_{3}$.
	
	The material (Fe$_{x}$Ni$_{1-x}$)$_{75}$P$_{16}$B$_{6}$Al$_{3}$ is an amorphous metal with 
	Ruderman-Kittel-Kasuya-Yosida 
	(RKKY) type of interactions between the magnetic ions. Its magnetic properties are mainly
	 determined by the Fe atoms since the magnetic moments of the Ni atoms are quenched due to 
	 charge transfer from the metalloids \cite{five}. For x=0, i.e. the pure Ni alloy, the system is 
	 non magnetic and with increasing Fe concentration  long range ferromagnetic order sets in 
	 at x=0.17. For $x < 0.17$ typical spin glass behaviour is found and for a range of concentrations
	  with $x > 0.17$, re-entrant spin glass behaviour occurs at low temperatures. The sample used in 
	  this investigation, (Fe$_{0.20}$Ni$_{0.80}$)$_{75}$P$_{16}$B$_{6}$Al$_{3}$ is in the form of a thin 
	  ribbon
	   (cross section 0.01*1.00 mm $^2$ and length 4 mm). The measurements were performed in a 
	   non-commercial superconducting quantum interference device (SQUID) 
	   magnetometer \cite{six} with 
	   the magnetic field applied along the length of the sample.

\section{Results and Discussion}

Fig. 1. shows ZFC, FC and TRM magnetisation vs. temperature for an applied field of 0.5 Oe. 
The onset of ferromagnetic long range order is signalled by a sharp increase of both the ZFC 
and the FC magnetisation just above the Curie temperature $T_c\approx 92$ K. A corresponding increase 
of the remanence is also observed, followed by a weak maximum and a continued increase at 
lower temperatures. A field dependent temperature, $T_{irr}(H)$,  where the magnetisation becomes 
irreversible can be defined at the temperature where the FC and ZFC curves merge. In the ZFC 
magnetisation a faint maximum is observed at $T\approx 20$ K. This signals the re-entrant spin glass
 phase that may establish at lower temperatures.

\begin{figure}

\centerline{\hbox{\epsfig{figure=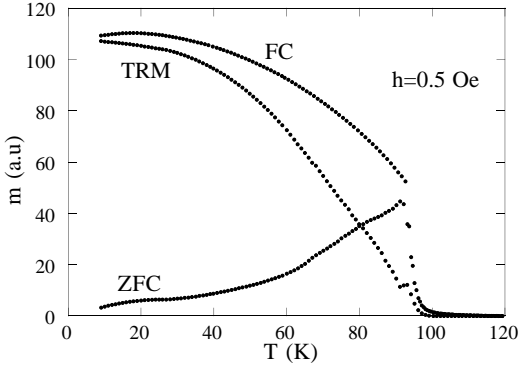,width=8.5cm}}}
\caption{
\hbox {The zero feld cooled (ZFC), field cooled (FC)} and thermo 
remanent magnetisation (TRM) contours in a field of $h$=0.5 Oe.
}
\label{fig1}
\end{figure}

	The FC magnetisation remains strongly magnetised throughout the FM phase and also into 
	the re-entrant spin glass 'phase' revealing that the aligned ferromagnetic domains 
	sustain even at low temperatures where the possible spin glass phase should appear. 
	This may be interpreted as if the ferromagnetic phase persists also in the low 
	temperature re-entrant spin glass 'phase' which may lead to the conclusion that no 
	true re-entrance exists in the material, but only a coexistence of spin glass and 
	ferromagnetic order \cite{seven}. On our experimental time scale this indeed seems to be the 
	case, but the crucial question to answer is which is the ground state. This question 
	can however not be answered from experimental observations on a non-equilibrium system
	 at one finite observation time. A dynamic scaling analysis intrinsically contains 
	 an extrapolation to infinite time scales and may thus allow some insight as to the
	  nature of the low temperature phase, such an analysis on the current system has 
	  been reported elsewhere \cite{two} and is indicative of the existence of a true re-entrant 
	  spin glass phase.
	
	In many disordered magnetic systems the response to an applied field is dependent on
	 the time the system has been kept at constant temperature in its 'glassy' state. 
	 
	 \begin{figure}

\centerline{\hbox{\epsfig{figure=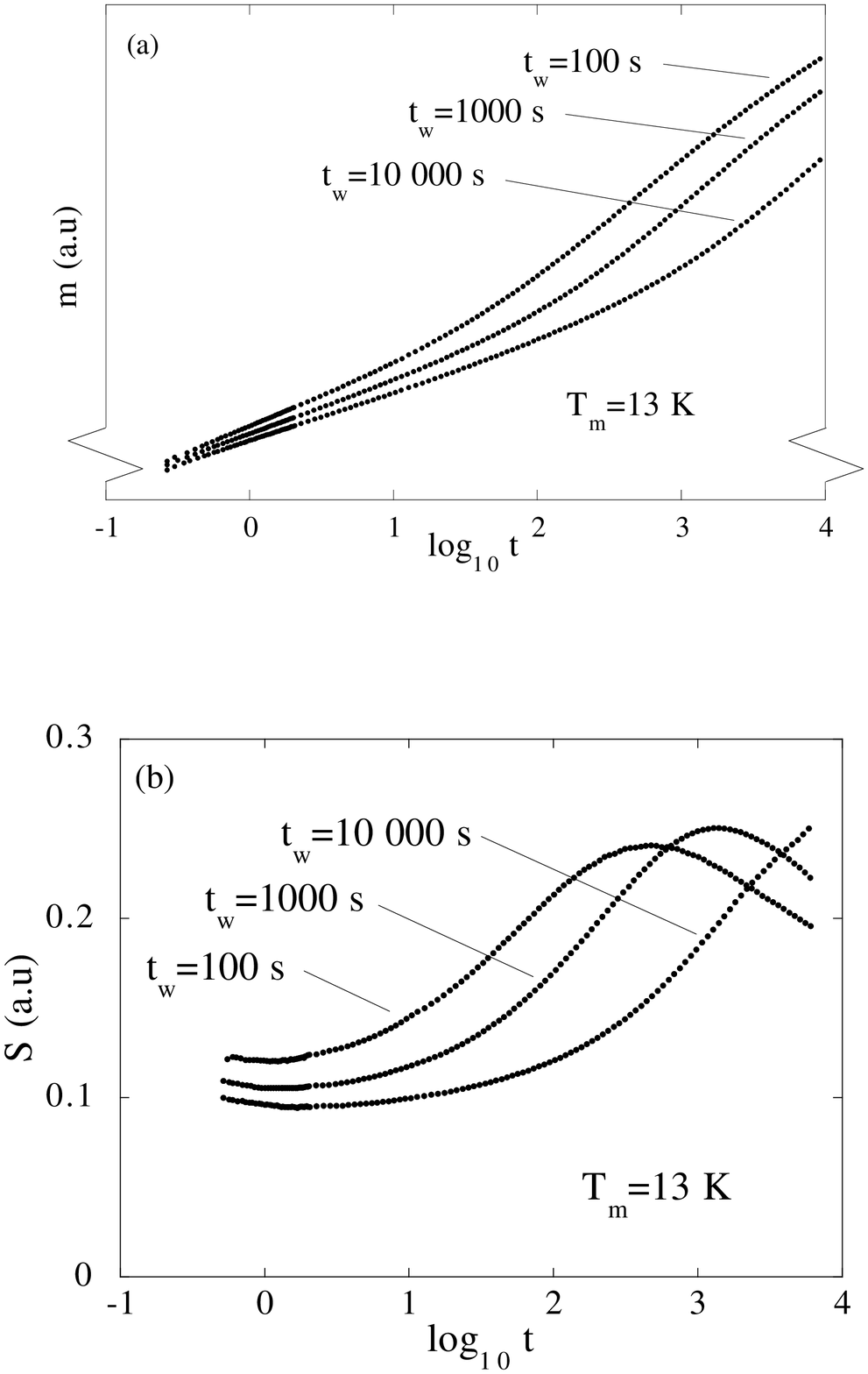,width=8.5cm}}}
\caption{
\hbox {$(m)$ vs. log$t$ (a) and corresponding relaxation rate}
$S(t)$=1/$h$ $dm/d$log$t$ vs. log$t$ (b), for $t_w=10^2, 10^3$ and 
$10^4$ s, at $T_m$=13 K and $h$=0.2 Oe. 
}
\label{fig2}
\end{figure}

	 This was first observed in spin glasses \cite{eight} but has later also been observed in 
	 various other disordered magnetic systems \cite{nine}. Such an ageing phenomenon, may e.g.
	  be disclosed by time-dependent zero-field-cooled (ZFC) magnetisation measurements 
	  at low enough fields. The sample is then cooled in zero field from a reference 
	  temperature, here 120 K, to a measurement temperature ($T_m$), and kept there a 
	  wait time ($t_w$). Thereafter a dc field ($h$) is applied and the magnetisation is
	   recorded as a function of time.
	   
	    In Fig. 2 (a) $m$ is plotted vs. 
	   log$t$ at
	    $T_m=13$ K, $h =0.2$ Oe  $t_w =10^2, 10^3$, and $ 10^4 $ s. The corresponding 
	    relaxation rate, $S(t)$ , 
	    is plotted in Fig. 2 (b). At this temperature the system is in its re-entrant 
	    spin glass 'phase' \cite{two}. A strong wait time dependence is observed. 
	    The m vs. log$t$ curve has an inflection point where log$t$ 
	    $\approx$ 
	    log$t_w$ and the relaxation
	     rate attains a corresponding maximum. A rather similar ageing behaviour is also 
	     observed in the ferromagnetic phase \cite{three}. The fact that ageing exists in both the
	      FM and the RSG regions may spontaneously be interpreted to suggest that the two 
	      phases are dynamically equivalent. However, if there exists a 
\begin{figure}
\centerline{\hbox{\epsfig{figure=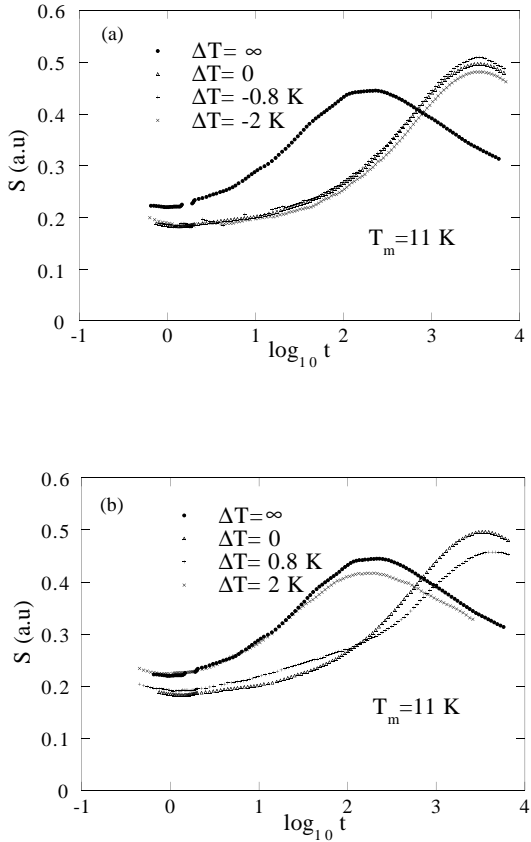,width=8.5cm}}}
\caption{
\hbox {The relaxation rates} $S(t)$=1/$h$ $dm/d$log$t$ 
for (a) negative $\Delta T$'s and (b) positive $\Delta T$'s
at $T_m$=11 K.
}
\label{fig3}
\end{figure}
	      true phase transition, 
	      the dynamic behaviour is expected to show some significant distinctions in-between 
	      the two phases. One striking change when passing from the RSG to the FM region is 
	      an enormous decrease of the magnitude of the magnetic field perturbation  that is 
	      needed to reinitialise an established ageing state \cite{four}. Here we discuss how an 
	      established ageing state is reinitialised due to a controlled temperature perturbation
	      , and again find striking differences in-between the behaviour in the RSG and the
	       FM phase.

	Disordered magnetic systems that display an ageing behaviour are affected by temperature
	 cyclings or shifts after the initial wait time \cite{ten}. This influence is clearly seen in 
	 the magnetic relaxation curves and confirm that the observed ageing effect originates 
	 from the chaotic nature of the underlying spin configuration. To examine the differences 
	 in the ferromagnetic and in the re-entrant spin glass phases temperature cycling
	  experiments were performed at two different temperatures, $T_{m}=40$
	  K and $T_m=11$ K.
	   The experimental procedure is as follows: first the system is cooled in zero field 
	   from the reference temperature, $T_{ref}=120$ K, to the measurement temperature. After a
	    wait time, $t_w=3000 $ s, the system is subjected to a temperature cycling of magnitude
	    , $\Delta T$, 
	    \begin{figure}
\centerline{\hbox{\epsfig{figure=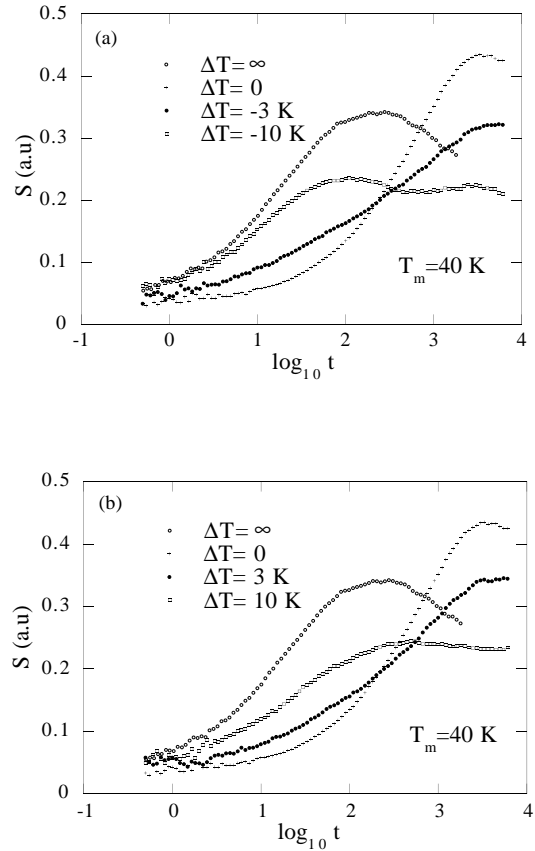,width=8.5cm}}}
\caption{
\hbox {The relaxation rates} $S(t)$=1/$h$ $dm/d$log$t$ 
for (a) negative $\Delta T$'s and (b) positive $\Delta T$'s
at $T_m$=40 K.
}
\label{fig4}
\end{figure}
\noindent a time $t_{\Delta T} =10$ s spent at the cycling temperature.

	    The time for heating
	     and cooling is not counted for in this measure. When $T_m$ is recovered the probing field
	      is applied, $h =0.2$ Oe, and the relaxation of the magnetisation is recorded as a 
	      function of time.

	Figure 3 (a) shows the relaxation rate $S (t)$ measured at $T_m=11$ K for two different
	 magnitudes of negative temperature cyclings, $\Delta T = -0.8$ K 
	 and $-2$ K. The curve marked 
	$\Delta T =\infty$ corresponds to a $\Delta T$ that takes the system above $T_c$. 
	This is the 'youngest' curve
	  that is possible to get by cooling directly to $T_m$ recording the relaxation without
	   wait time. The achieved maximum at $t \approx 100$ s is assigned to an effective wait time 
	   governed by the cooling rate and the time allowed for temperature stabilisation 
	   when $T_m$ is recovered. The $\Delta T$ =0 curve is the relaxation rate of an ordinary ZFC
	    relaxation curve measured with $t_w =3000$ s. The two negative temperature cycles 
	    have a very weak influence on the relaxation rate. The magnitude of the maximum 
	    at $ t \approx 3000$ s is slightly altered compared to the $\Delta 
	    T =0$ K  curve. 
	    In Fig. 3 (b) the result for two positive temperature cyclings with the 
	    same magnitude as in the previous figure is presented. The difference is striking.
	     The maximum in $S(t) $ at$ t \approx 3000$ s decreases and for $\Delta 
	     T= 2$ K the systems becomes
	     totally reinitialised, the maximum occurs at shorter times, almost 
	      coinciding with the $\Delta T=\infty $ behaviour. This means that a positive temperature
	       cycle of 2 K, has almost the same effect as cooling the system 
	       directly to $T_m= 11$ K.
	        A similarly asymmetric response for  positive and negative $\Delta T$ 's is observed in
	         ordinary spin glasses \cite{ten}.

	In Figure 4 the relaxation rate is displayed for $T_m=40$ K, in the ferromagnetic phase, 
	for negative (a) and positive (b) $\Delta T$ 's. The magnitude of the temperature cyclings are
	 chosen so  $\Delta T/T_m$  is comparable at the two different measurement temperatures. 
	 Here, both the positive and negative temperature cycles reinitialise the system
	  an equal amount. The response is symmetric,  positive and negative $\Delta T$ 's of equal
	   magnitude have qualitatively the same  effect on the relaxation rate in the 
	   ferromagnetic phase. This is in contrast with the response in the RSG phase, 
	   where the result on the relaxation rate is highly asymmetric for different signs
	    on the $\Delta T$ 's. Another notable difference in the two phases is that it is easier 
	    to reinitialise the system with a positive $\Delta T$  in the RSG phase than in the FM phase. 
	    The RSG phase is stable' against negative $\Delta T$ 's but extremely sensitive to 
	    positive $\Delta T$ 's.
	     The ferromagnetic phase is equally stable for positive and negative $\Delta T$ 's and
	      is more difficult to reinitialise  completely.

\section{Conclusions}

	We have studied some dynamic and static properties of the re-entrant ferromagnet
	  (Fe$_{0.20}$Ni$_{0.80}$)$_{75}$P$_{16}$B$_{6}$Al$_{3}$. 
	  The ZFC-FC protocol results in an irreversibility 
	 temperature, Tirr, that coincides with the onset of long range ferromagnetic 
	 order, $T_c\approx 92$ K. The FC magnetisation remains highly magnetised throughout the 
	 FM phase and into the RSG phase revealing that ferromagnetic domains are 
	 present even at low temperatures at our experimental time scales. 
	
	The relaxation rate, $S (t)$, displays distinct differences for the FM and the 
	RSG regions when the system is exposed to a temperature cycle prior the field 
	application. In the RSG 'phase' the system behaves like an ordinary SG with an 
	asymmetric result for negative and positive $\Delta T$ 's. The underlying spin configuration 
	is very unstable against positive $\Delta T$ 's but looks stable against negative $\Delta T$ 's,
	 if 
	the duration of the cycle is constant and 'short', $t_{\Delta T} 
	\approx  10$ s. In the FM region, the 
	result for $\pm$ $\Delta T$ 's are symmetric. The ferromagnetic phase  is equally  stable to
	 positive and negative $\Delta T$' s. The FM phase is harder to completely reinitialise 
	 than the 'RSG' phase if a positive $\Delta T$ is used. The fact that there is a 
	 re-initialisation  (only) when large enough temperature perturbations are used,  
	 support that there is chaos between the states attained at different temperatures
	  and that there is an overlap between states . One conclusion to be drawn from 
	  these results is that the spin glass phase has barrier heights for domain 
	  growth that increases with decreasing temperature, whereas in the ferromagnetic 
	  phase the domain growth is governed by thermal activation over barriers of 
	  constant height.

\section{Acknowledgement}

Financial support from the Swedish Natural Science Research Council (NFR) 
is gratefully acknowledged.

\end{multicols} 

\end{document}